\documentclass{vgtc}                          

\ifpdf
  \pdfoutput=1\relax                   
  \usepackage{graphicx}                
  \DeclareGraphicsExtensions{.pdf,.png,.jpg,.jpeg} 
\else
  \usepackage{graphicx}                
  \DeclareGraphicsExtensions{.eps}     
\fi%

\graphicspath{{figures/}{pictures/}{images/}{./}} 

\usepackage{microtype}                 
\PassOptionsToPackage{warn}{textcomp}  
\usepackage{textcomp}                  
\usepackage{mathptmx}                  
\usepackage{times}                     
\usepackage{cite}                      
\usepackage{tabu}                      
\usepackage{booktabs}                  
\usepackage{expdlist}
\newcommand{\inlinevis}[3]{\raisebox{#1}[0pt][0pt]{\includegraphics[height=#2]{#3}}}
\newcommand{\tp}[1]{\includegraphics[width=\picturewidth]{#1}}

\newcommand{\icon}{\texttt{Icon Only}\,\inlinevis{-2pt}{1.2em}{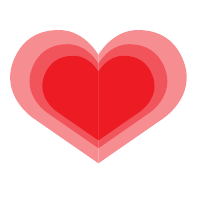}}
\newcommand{\icontext}{\texttt{Icon+Text}\,\inlinevis{-3pt}{1.3em}{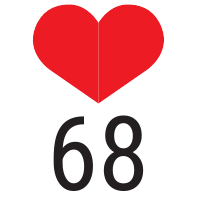}}
\newcommand{\chartonly}{\texttt{Chart Only}\,\inlinevis{-3pt}{1.3em}{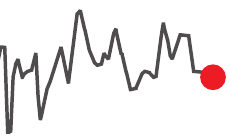}}

\newcommand{\charttext}{\texttt{Chart+Text}\,\inlinevis{-3pt}{1.3em}{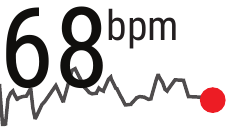}}

\newcommand{\textonly}{\texttt{ Text Only}\,\inlinevis{-3pt}{1.3em}{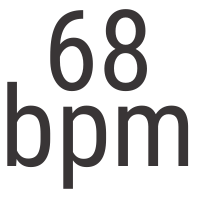}}

\usepackage[T1]{fontenc}
\usepackage{graphicx,calc}

\newcommand{\bpstart}[1]{\vspace{1mm} \noindent{\textbf{#1.}}}

\newcommand{\health}[1]{\textcolor[rgb]{0,.8,.59}{#1}}
\newcommand{\weather}[1]{\textcolor[rgb]{0.2,.6,1}{#1}}
\newcommand{\device}[1]{\textcolor[rgb]{.72,.47,.26}{#1}}

\newcommand{\changes}[1]{\textcolor[RGB]{0,0,0}{#1}}

\newlength{\picturewidth}

\onlineid{0}

\vgtccategory{Research}

\vgtcinsertpkg

\title{Visualizing Information on Watch Faces:
A Survey with Smartwatch Users}

\author{Alaul Islam\thanks{e-mail: mohammad-alaul.islam\,|\,petra.isenberg@inria.fr}\\ %
        \scriptsize Université Paris-Saclay, CNRS, Inria, LRI %
\and Anastasia Bezerianos\thanks{e-mail: anastasia.bezerianos@lri.fr}\\ %
     \scriptsize Université Paris-Saclay, CNRS, Inria, LRI %
\and Bongshin Lee\thanks{e-mail: bongshin@microsoft.com}\\ %
     \scriptsize Microsoft Research%
\and Tanja Blascheck\thanks{e-mail: tanja.blascheck@vis.uni-stuttgart.de}\\
    \scriptsize University of Stuttgart\\
\and Petra Isenberg\textsuperscript{*}\\
\scriptsize Université Paris-Saclay, CNRS, Inria, LRI
}

\teaser{
  \centering
  \includegraphics[width=\linewidth]{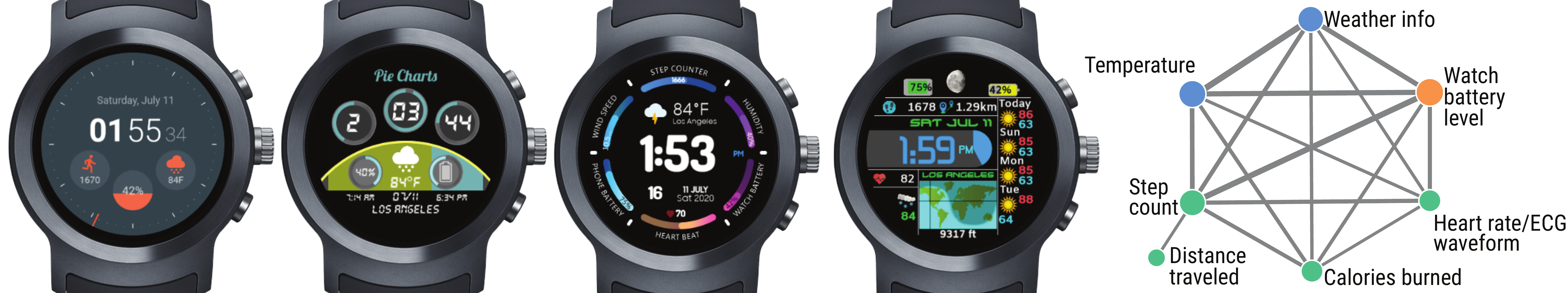}
  \caption{Smartwatch face examples (from Facer \cite{Facer:2020}) with increasing amounts of data items and representation types. From left to right: Material Volcano (BlueIceshard), Pie Charts II (Sunny Liao), Minimal Colors H (AK Watch), and Earthshade (Brad C). The graph on the right shows common pairs of data types displayed on the watch faces our 237 survey participants used. Circle colors correspond to three data categories: \health{Health \& Fitness}, \weather{Weather \& Planetary}, and \device{Device \& Location}.}
  \label{fig:teaser}
}

\abstract{
People increasingly wear smartwatches that can track a wide variety of data. However, it is currently unknown which data people consume and how it is visualized. To better ground research on smartwatch visualization, it is important to understand the current use of these representation types on smartwatches, and to identify missed visualization opportunities. We present the findings of a survey with 237 smartwatch wearers, and assess the types of data and representations commonly displayed on watch faces. We found a predominant display of health \& fitness data, with icons accompanied by text being the most frequent representation type. Combining these results with a further analysis of online searches of watch faces and the data tracked on smartwatches that are not commonly visualized, we discuss opportunities for visualization research. Supplementary material is available at \href{https://osf.io/nwy2r/}{https://osf.io/nwy2r/}.
} 

\CCScatlist{
  \CCScatTwelve{Human-centered computing}{Visu\-al\-iza\-tion}{Empirical studies in visualization}{};
  \CCScatTwelve{Human-centered computing}{Mobile devices}{}{}
}

\begin{document}
\maketitle

\section{Introduction}
According to research and market reports, the demand for smartwatches is expected to rise at a Compound Annual Growth Rate (CAGR) of 14.5\% between 2020 and 2025 \cite{RAM:2020}. People already use smartwatches as personal data collection devices, and with additional wifi connectivity they have access to various types of data. \changes{Smartwatches that use visualizations to display data and expose patterns, trends, or outliers in a compact way and at a glance may have many potential benefits.}
%
Yet, the small display size of smartwatches also creates unique challenges \cite{8443125} that call for visualization research. 

For device-oriented research, it is important to understand current use and practices of its adopters. Thus, in this work, we investigate the use of visualizations on watch faces, which are the first screen or home screen wearers see when glancing at or turning on their watch \cite{Zhang:2019:EnergyInefficienciesWatchFaces, Apple:2020}. These watch faces \changes{are typically small, 
have a resolution between
128--480 px per side with a viewable area of around 30--40mm \cite{8443125}
and} 
show the current time together with several  data types, such as step count, location, and weather information. Watch faces are often customizable, allowing wearers to choose the data they want to see regularly and at a glance. 
\changes{Given the large variety of data available to display on smartwatches, we were particularly interested to answer }
the following research questions: 
\begin{description}[\compact\setlabelphantom{Q1:}\itshape]
\item[Q1:] Which data types do people show on their watch faces?
\item[Q2:] In which form is the data currently represented? 
\item[Q3:] What more can we visualize? 
\end{description}

To answer these questions, we first conducted an online survey with smartwatch wearers, then complemented these results with an online search and analysis of smartwatch face examples, as well as an analysis of the technical capabilities of the watches our participants reported wearing. 
%
We contribute findings of current smartwatch use and open opportunities for visualization research and design. 
%

\section{Related Work}

\bpstart{Smartwatch use}
Similar to our research goal, several prior studies investigated smartwatch use in the wild. Schirra and Bentley  \cite{Schirra:2015:UEUCSW} and Cecchinato et al.\  \cite{Cecchinato:2017:UESR} conducted interviews with early adopters of smartwatches with a focus on why watches were adopted and what tasks they were used for. 
%
%
%
Later studies focused at commonly used features of smartwatches, finding that people mainly used smartwatches to monitor and track activities or respond to notifications in addition to timekeeping \cite{10.1145/3025453.3025993, 10.1145/2858036.2858522, 8374228, 7457170}. 
%
Others looked at specific smartwatch use such as in classrooms \cite{10.1145/2851581.2892493}, 
for healthcare purposes\cite{kamivsalic2018sensors, 10.1145/3341162.3346276, 7299348, 8880821, 10.1145/3123024.3125614}, stress detection \cite{10.1145/3341162.3344831}, real-time eating activity detection \cite{10.5555/3400306.3400354}, or understanding a wearer's emotional state \cite{10.1145/3123024.3125614}.
%
%
In contrast, we are not interested in particular applications or feature use. Instead we focus on \emph{what} information is displayed directly on watch faces, and \emph{how} it is displayed, outside of any particular watch app.

\bpstart{Visualizations on smartwatches}
Research on smartwatches in visualization is still sparse. The few publications that do exist focused either on studying representations for smartwatches or on designing representations for these small displays. For example, researchers studied low-level perceptual tasks to understand glanceability of smartwatch visualizations  \cite{8443125}, the impact of visual parameters (e.g., size, frequency, and color) on reaction times \cite{10.1145/2935334.2935344}, or representation preferences in an air traffic control use case \cite{Neis:2016}.

Others' visualization research described novel visualization designs specifically for smartwatches. Examples include research on representing health and fitness data on smartwatches \cite{10.1145/3154862.3154879,PMID:30741218}, for line charts \cite{Neshati:2019:10.20380/GI2019.23}, temporal data \cite{e2da2cc89b474b6ebbf3817b4bb025c6}, activity tracking more broadly \cite{10.1145/2971648.2971754}, and even for integrating visualizations in watch straps \cite{10.1145/3313831.3376199}.
In contrast to these works, our study contributes information on people's current representation types on watch faces and the results can be used to inform future research such as reported above. 


\section{Methodology}
We conducted an anonymous online survey, for which we recruited regular smartwatch wearers at least 18 years of age.\footnote{IRB approved under ref.\ no Paris-Saclay-2020-002 CER.}

\bpstart{Survey design}
Our survey consisted of three sections, primarily containing close-ended questions. 
The first was designed to elicit general information about a respondent's watch face. Here, we asked questions about the respondent's watch shape and in which form (analog, digital, or both analog \& digital) they read the time \changes{on the first screen or home screen of their watch}. The second section focused on which additional data types---such as step count or temperature---were shown on the respondent's watch face. 
In addition to offering common kinds of data types as options, we had an \emph{other} text field for participants to fill out in case their watch face showed data not in our list. 

\begin{figure}[tb]
\centering
  \includegraphics[width=\columnwidth]{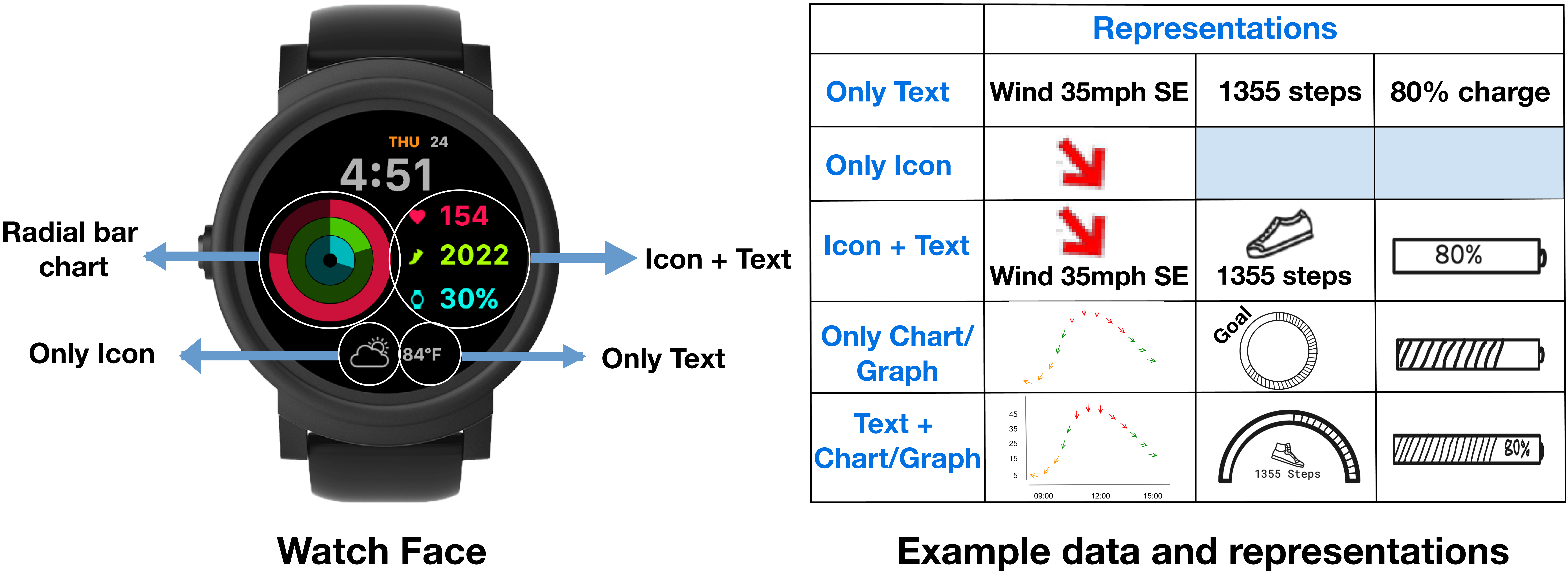}
  \vspace{-6mm}
  \caption{Explanatory image of answer choices shown to participants.} 
  \vspace{-3mm}
  \label{fig:explanatoryImage}
\end{figure}

To derive the list of data types for our survey (\autoref{tab:types-categories}), we consulted prior research \cite{10.1145/3025453.3025817} and analyzed images of popular watch faces from Facer \cite{Facer:2020}, a watch-face download and generation page/app for Android, Samsung, and iOS watches. Inspired by categories used in the Facer app, we grouped possible kinds of data into three categories: \health{health \& fitness} related data, \weather{weather \& planetary} data, and \device{device- \& location-related} data. 
\begin{table}[t]
\caption{Categories of data types shown on watch faces.}
\renewcommand{\arraystretch}{1.2}    
    \centering
    \small
    \begin{tabular}{p{2cm}|p{5.6cm}}
    \toprule
        \textbf{Category} & \textbf{Data type} \\ 
        \toprule
        \health{Health \& Fitness} & Heart rate/ECG waveform, step count, sleep related info (e.g., quality, duration), distance traveled, calories burned, floors/stairs climbed, and blood pressure \\ \hline
        \weather{Weather \& Planetary} & Weather info (e.g., sky condition), temperature, wind speed/direction, moon phase, humidity, and sunset/sunrise time \\ \hline
        \device{Device \& Location} & Watch battery level, phone battery level, bluetooth, wifi, and location name \\ \hline
        Other & Data and representation type not in our list (open textfield) \\ 
    \bottomrule
    \end{tabular}
    \vspace{-3mm}
    \label{tab:types-categories}
\end{table}
%
%
%
%
%
For each kind of data we asked participants to tell us \emph{how} the data was shown on their watch face. We provided participants with five possible representation types accompanied by a text description (\autoref{tab:representation-categories}) and by an explanatory image (\autoref{fig:explanatoryImage}). 
These categories were based on how numerical or categorical data are displayed on more than 500 watch faces that we collected from the Facer app and internet searches. 

\begin{table}[t]
\caption{Representation types on watch faces.}
\renewcommand{\arraystretch}{1.2}    
    \centering
    \small
    \begin{tabular}{p{1.6cm}|p{6.0cm}}
    \toprule
        \textbf{Representation 
        } & \textbf{How data is displayed} \\
        \toprule
        Only Text & as text, including numbers (e.g., text to display heart rate \inlinevis{-3pt}{1.3em}{images/text.pdf}) \\ \hline
        Only Icon & as an icon (e.g., a pulsating heart representing heart rate \inlinevis{-2pt}{1.2em}{images/icon.pdf}) \\ \hline
        Icon + Text & as text with an icon for context (e.\,g.,\ a static heart with text to show the current heart rate \inlinevis{-3pt}{1.3em}{images/icontext.pdf}) \\ \hline
        Only Chart/Graph & as a simple statistical chart (e.\,g.,\ showing recent heart rates \inlinevis{-3pt}{1.3em}{images/chartonly.pdf}) \\ \hline
        Text + Chart/Graph & as text with a simple chart (e.g., heart rate linechart \inlinevis{-3pt}{1.3em}{images/charttext.pdf}) \\
    \bottomrule
    \end{tabular}
    \label{tab:representation-categories}
\end{table}



In the final section of the survey we asked participants to provide the brand and model name for their smartwatch so we could verify the plausibility of their responses. We also asked participants to optionally upload a picture or screenshot of their watch face for verification. More details about the questions and format are available in the supplementary material.


\bpstart{Participant recruitment}
To reach a wide range of smartwatch wearers we advertised our survey on popular social media (Reddit, Twitter, Facebook, Instagram, and LinkedIn), and asked colleagues to spread the call to their labs. The survey was available online for 30 days during April and May, 2020.

\bpstart{Data quality}
We took several steps to ensure the quality of our collected data. 
From the 463 total responses, 177 were incomplete and another 31 failed our screening procedure. 
We asked participants to wear a smartwatch or at least have it available around them (e.g., charging, holding) to ensure that they do not answer questions from memory. We prompted them to verify if that was the case. The 30 participants who answered ``no'' were not allowed to continue to the survey. We also excluded one participant who did not sign the consent form. We had 255 complete responses for data analysis.
We discarded 18 additional participants: Five of them reported to seeing every single kind of data, and their responses did not match the watch face image they provided. Three participants reported the names of several smartwatch models, so we could not determine which one they recorded during the study. Another 10 wore fitness bands rather than smartwatches and were excluded due to their dedicated focus on fitness data and limited display capabilities. We report results from the remaining 237 valid responses.
%
%

\section{Analysis \& Results}

The majority of participants reported wearing a smartwatch with a round display (150\texttimes), followed by a square (68\texttimes), and rectangular display (17\texttimes). Two participants reported having Squaricle / Rounded square types. 
Most participants (149\texttimes) reported that the data items on their watch face are static and do not change (automatically or manually, e.\,g.,\, by tap or swipe). Forty six participants reported their watch face changed automatically while 42 reported that they could manually swap data shown on their watch face. Participants' smartwatches came from 20 different brands with
Apple~(76\texttimes), Fossil~(51\texttimes), Samsung~(36\texttimes), Garmin~(17\texttimes), and Huawei~(12\texttimes) being the top five brand (80\% of our respondents). 


\begin{figure}[tb]
\centering
  \includegraphics[width=.8\columnwidth]{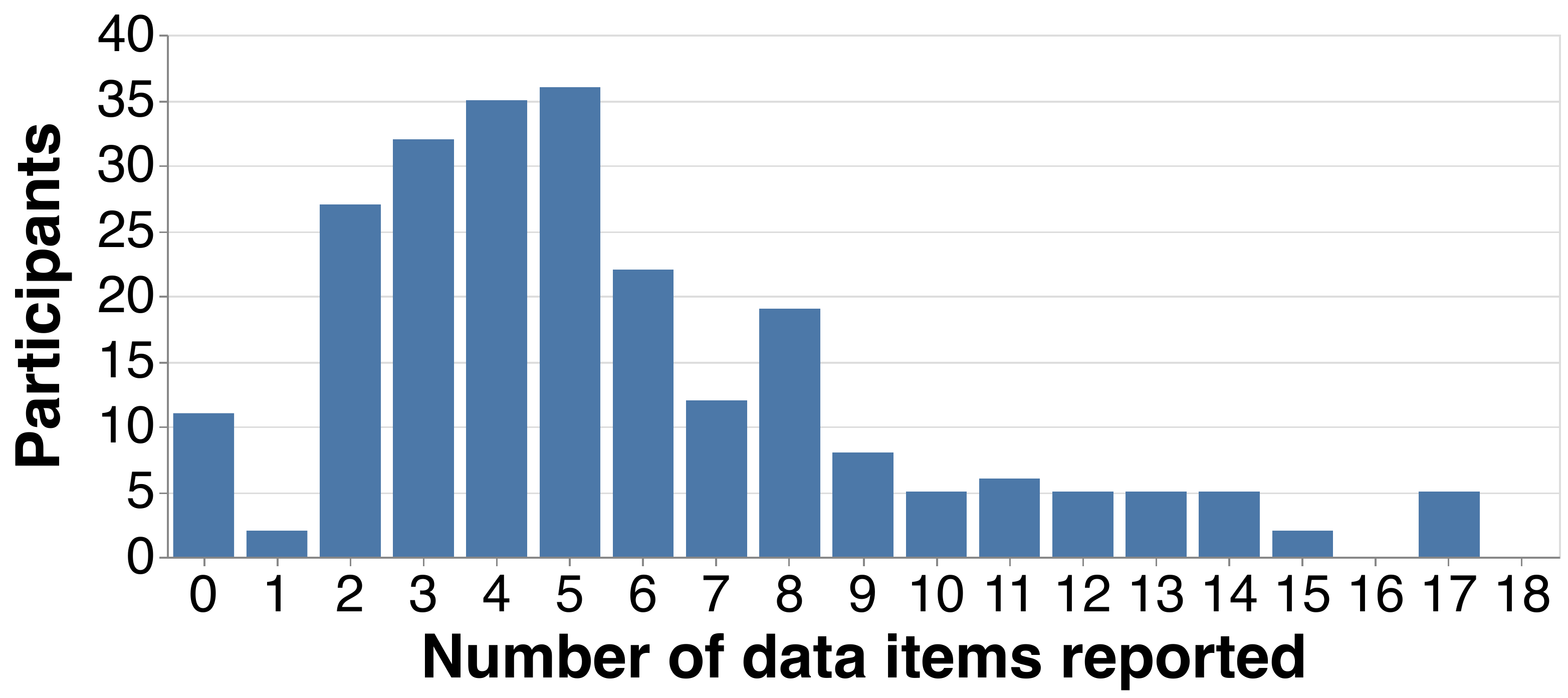}
 \vspace{-3mm}
  \caption{Number of data items present on a respondent's watch face.}
\vspace{-3mm}
  \label{fig:datatypesReported}
\end{figure}

\begin{figure}[t]
\centering
\includegraphics[width=\columnwidth]{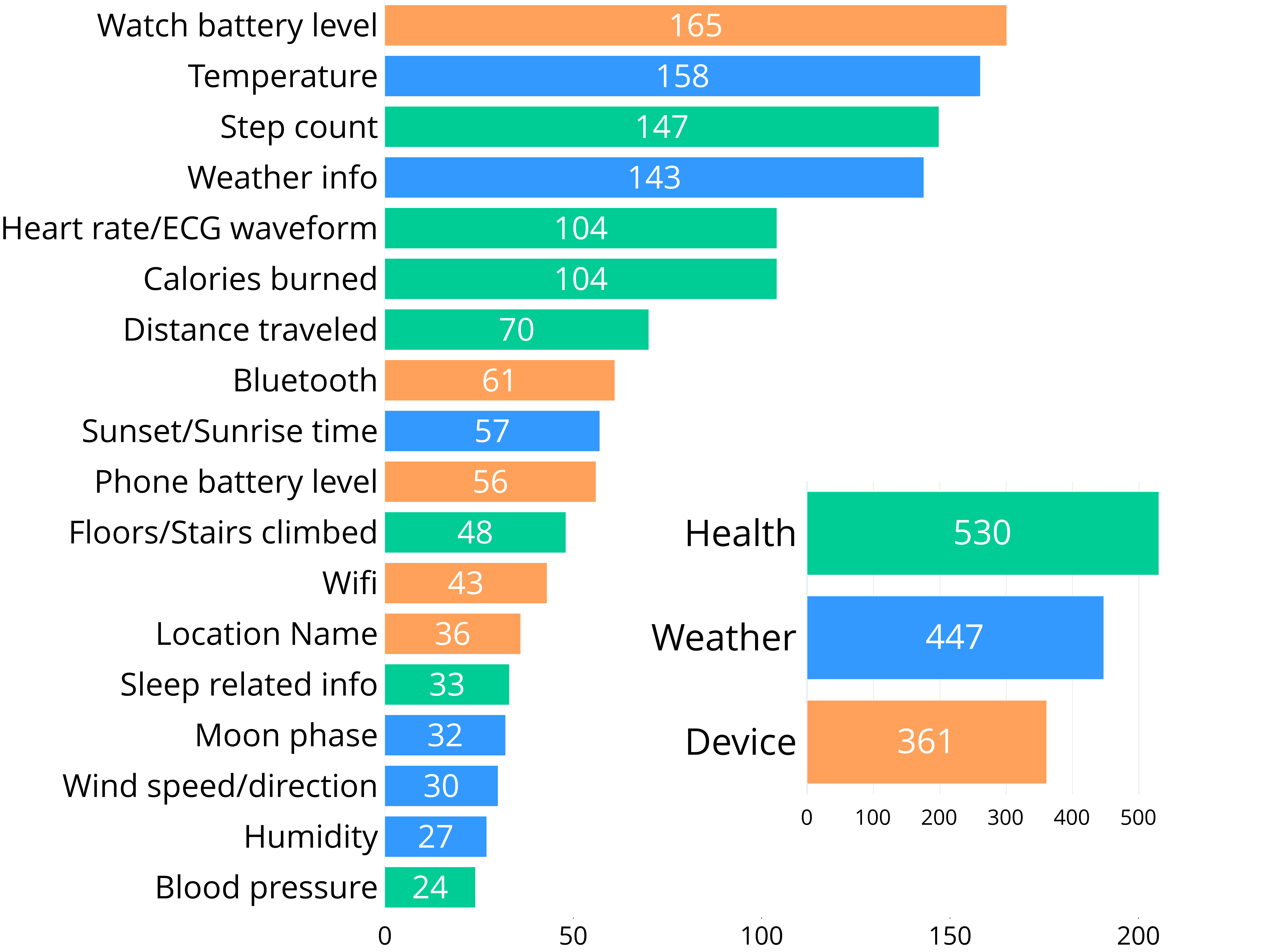}
\vspace{-7mm}
\caption{Distribution of data types participants displayed and saw on their watch faces (left); aggregated by categories on the right.}
\vspace{-3mm}
  \label{fig:mostDisplaedData}
\end{figure}


\subsection{Q1: Which data types do people show on their watch faces?}

We were first interested to see whether people had configured their watch faces to show a large amount or only a few data items. On average, participants reported showing a 
median of 5 different data items on their watch faces. 
\autoref{fig:datatypesReported} shows that having 3, 4, or 5 data items were the most common answers.

Next, we wanted to learn which data types were the most commonly displayed (\autoref{fig:mostDisplaedData}). From the three categories we asked about, 
\health{health-fitness} related data were the most commonly reported (530\texttimes). 
%
%
The most common data type in this category was \emph{step count} (the third most common overall, 147\texttimes). \emph{Temperature} was the most frequent \weather{weather \& planetary} data type (the second most common overall, 158\texttimes). For \device{device-location} related data, watch \emph{battery level }(165\texttimes) was the most displayed and also the most common overall. 
The most commonly mentioned data types from the free-text responses were: \textit{standing up count} (43\texttimes) and \textit{exercise/body movement time} (24\texttimes).

Next, we wanted to learn about individual watch faces. We analyzed, which categories were most common per watch face and which data types often appeared together. On average, most of the data shown on an individual watch face came from the \health{health \& fitness} category. Participants reported seeing on average: 2.24 
\health{health \& fitness} (\textit{Mdn} = 2, 95\% CI [1.98, 2.48]), 1.89 
\weather{weather \& planetary} (\textit{Mdn} = 2, 95\% CI [1.69, 2.08]) and 1.52 
\device{device-location} related data (\textit{Mdn} = 1, 95\% CI [1.35, 1.7]) on their watch face.



To know more about which types of data are commonly shown together, we performed a co-occurrence analysis of data types participants saw on their watch faces. The graph in \autoref{fig:teaser} shows combinations of two kinds of data that can be found on at least 25\% of our respondents' watch faces. The thicker the link, the more frequent the data pair appeared on people's watch faces. Circle size corresponds to how often participants reported seeing this data type. Circle color corresponds to the data type category. Only connections that appeared more than 59 (${\approx }$ 237 / 4) times are shown. 

\subsection{Q2: In which form is the data currently represented?}
\begin{figure}[t]
\centering
  \includegraphics[width=\linewidth]{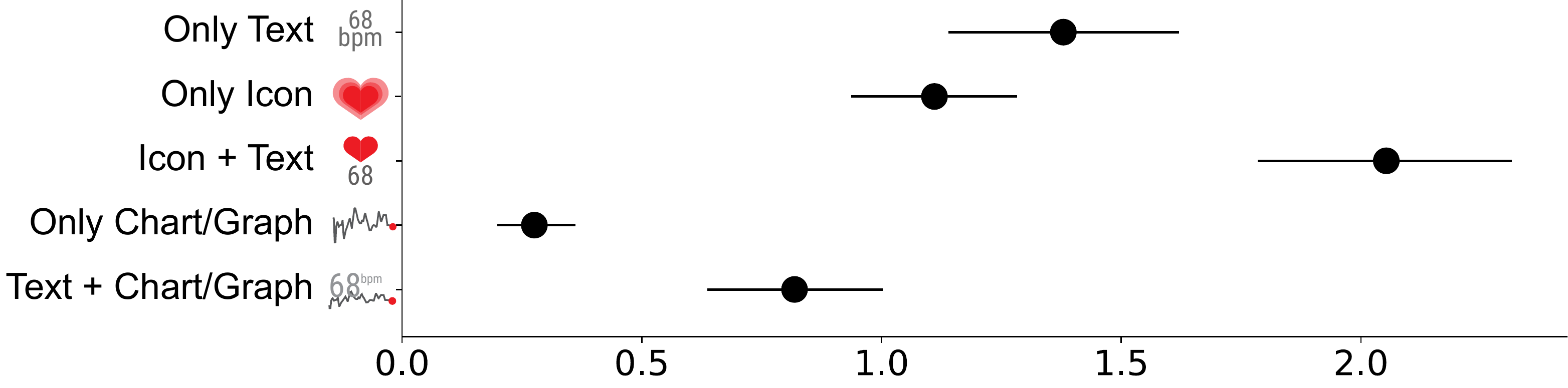}
  \vspace{-7mm}
\caption{Average number of representation types for each participant. }
  \label{fig:avgDataRepresentation}
\end{figure}

\autoref{fig:avgDataRepresentation} shows the average number of representation types each participant had on their watch face. 
\icontext\ was the most common representation type, used to display on average two kinds of data types on each watch face (\textit{M} = 2.05, 95\% CI: [1.78, 2.32]). The next most common were \textonly\ (\textit{M} = 1.38, 95\% CI: [1.13, 1.66]), and \icon\ (\textit{M} = 1.11, 95\% CI: [0.93, 1.3]). Representations using visualizations were less common. \charttext\ (\textit{M} = 0.82, 95\% CI: [0.64, 1.03]) and \chartonly\ (\textit{M} = 0.28, 95\% CI: [0.2 , 0.37]) appeared less than once per watch face on average. 

\begin{figure}[t]
\centering
  \includegraphics[width=.93\columnwidth]{images/data_vs_representation_reported.pdf}
  \vspace{-3mm}
\caption{Representation types reported for different data types.}
\vspace{-3mm}
  \label{fig:dataVsDisplayReported}
\end{figure}




In \autoref{fig:dataVsDisplayReported} we can see how many participants showed each data type with each representation type. Data types most commonly displayed with either \chartonly\ or \charttext\ were \emph{calories burned} ($14+30=44$\texttimes), \emph{step count} ($10+32=42$\texttimes), and \emph{watch battery levels} ($14+28=42$\texttimes). 


\bpstart{Complementary search of representation types}
Surprised by the high number of icons reported,
we decided to investigate further how different information can be displayed on watch faces. 
We conducted an extensive image search,
during which we looked for examples of each representation type in current use.  We looked at popular watch brands' websites, searched the internet for images (keywords: smartwatch face, popular smartwatch, smartwatch, etc.), and looked at examples from the Facer watch face creation and distribution app. \autoref{tab:exampleimages} shows exemplary graphics for each kind of data \texttimes\ representation type combination, redrawn for image clarity. 
%
%
We found only few examples online of data types represented by an \icon\ display. Yet, \autoref{fig:dataVsDisplayReported} shows that participants reported seeing \icon\ representations for almost every data with on average around one \icon\ display per smartwatch face. We discuss this discrepancy further in \autoref{sec:discussion}.
%
%

\setlength{\picturewidth}{.115\columnwidth}
\newcolumntype{C}[1]{>{\centering\arraybackslash}m{#1}}

\begin{table}[t]
        \caption{Redrawn example representations from real smartwatch faces. Text color corresponds to the data type category. Bluetooth and wifi only text and only icon change color based on on/off status.}
    \renewcommand{\arraystretch}{1.2}    
    \centering
    \footnotesize
    \begin{tabular}{@{}p{.27\columnwidth}|@{}C{.13\columnwidth}@{}|@{}C{.13\columnwidth}@{}|@{}C{.13\columnwidth}@{}|@{}C{.13\columnwidth}@{}|@{}C{.13\columnwidth}}
    \toprule
        \textbf{Data Types} & \textbf{Only Text}&\textbf{Only Icon}&\textbf{Icon+\-Text}&\textbf{Only Chart}&\textbf{Text+\-Chart}\\
        \toprule
        \health{Heart rate / ECG waveform} & \tp{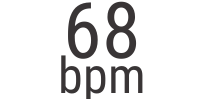}
        &\tp{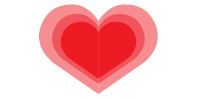}&\tp{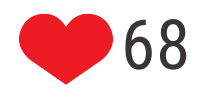}&\tp{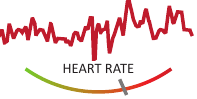}&\tp{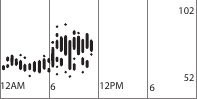}\\
        \health{Step count}&\tp{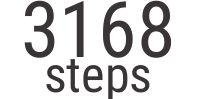}& &\tp{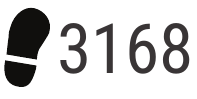}&\tp{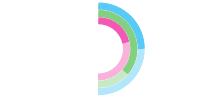}&\tp{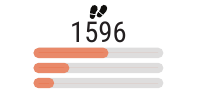}\\
        \health{Sleep related info}&\tp{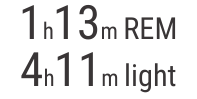}& &\tp{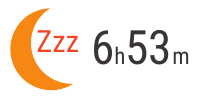}&\tp{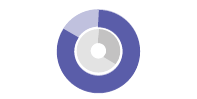}&\tp{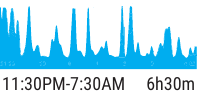}\\
        \health{Distance traveled}&\tp{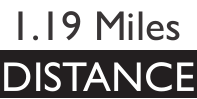}&&\tp{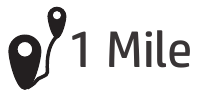}&\tp{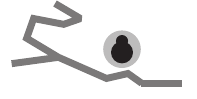}&\tp{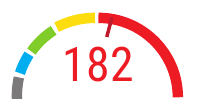}\\
        \health{Calories burned}&\tp{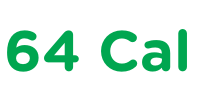}&&\tp{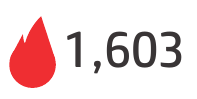}&\tp{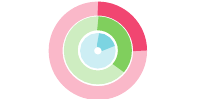}&\tp{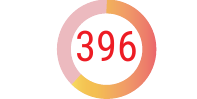}\\
        \health{Floors/Stairs climbed}&\tp{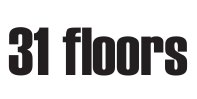}&&\tp{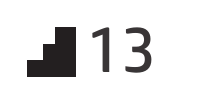}&\tp{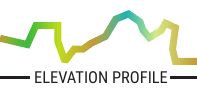}&\tp{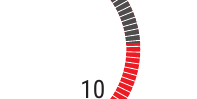}\\
        \health{Blood pressure}&\tp{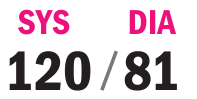}&&\tp{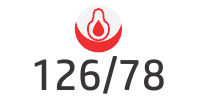}&&\tp{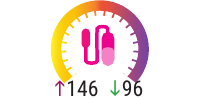}\\
        \hline
        \weather{Weather info}&\tp{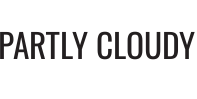}&\tp{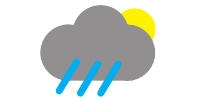}&\tp{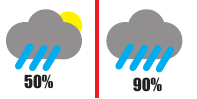}&&\tp{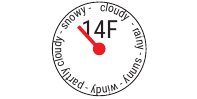}\\
        \weather{Wind speed/direction}&\tp{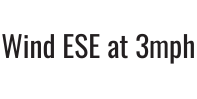}&&\tp{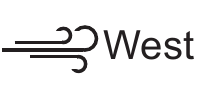}&&\tp{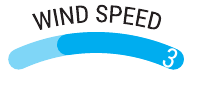}\\
        \weather{Temperature}&\tp{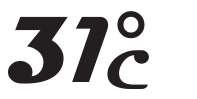}&&\tp{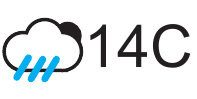}&&\tp{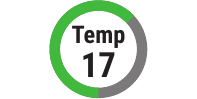}\\
        \weather{Sunset/Sunrise time}&\tp{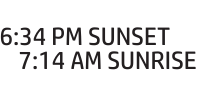}&\tp{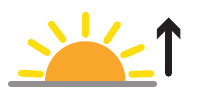}&\tp{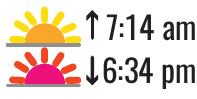}&&\tp{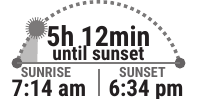}\\
        \weather{Moon phase}&\tp{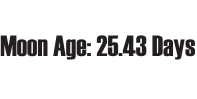}&\tp{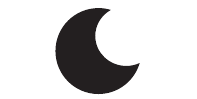}&\tp{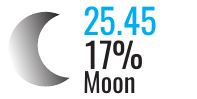}&\tp{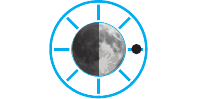}&\tp{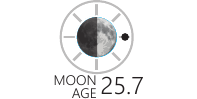}\\
        \weather{Humidity}&\tp{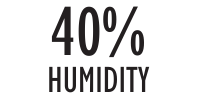}&&\tp{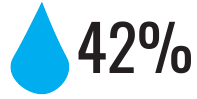}&&\tp{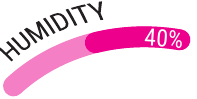}\\
        \hline
        \device{Bluetooth}&\tp{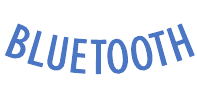}&\tp{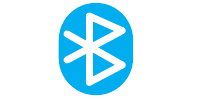}&&\tp{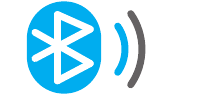}&\\
        \device{Phone battery level}&\tp{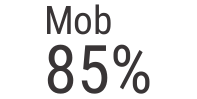}&&\tp{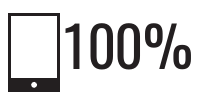}&\tp{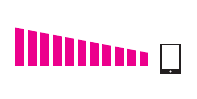}&\tp{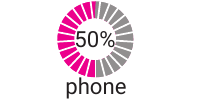}\\
        \device{Location name}&\tp{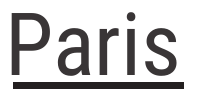}&&\tp{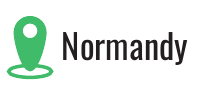}&\tp{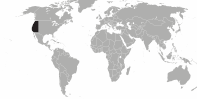}&\tp{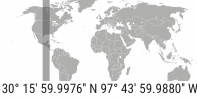}\\
        \device{Wifi}&\tp{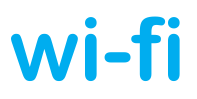}&\tp{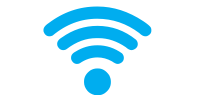}&\tp{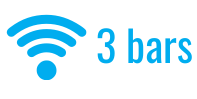}&\tp{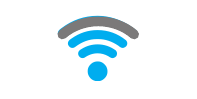}&\tp{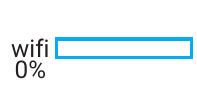}\\
        \device{Watch battery level}&\tp{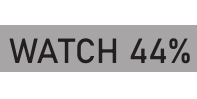}&&\tp{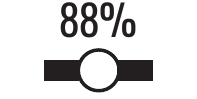}&\tp{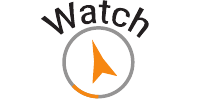}&\tp{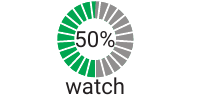}\\
    \bottomrule
    \end{tabular}
    \vspace{-3mm}
    \label{tab:exampleimages}
\end{table}




\subsection{Q3: What more can we visualize?}
\begin{figure}[tb]
\centering
  \includegraphics[width=\columnwidth]{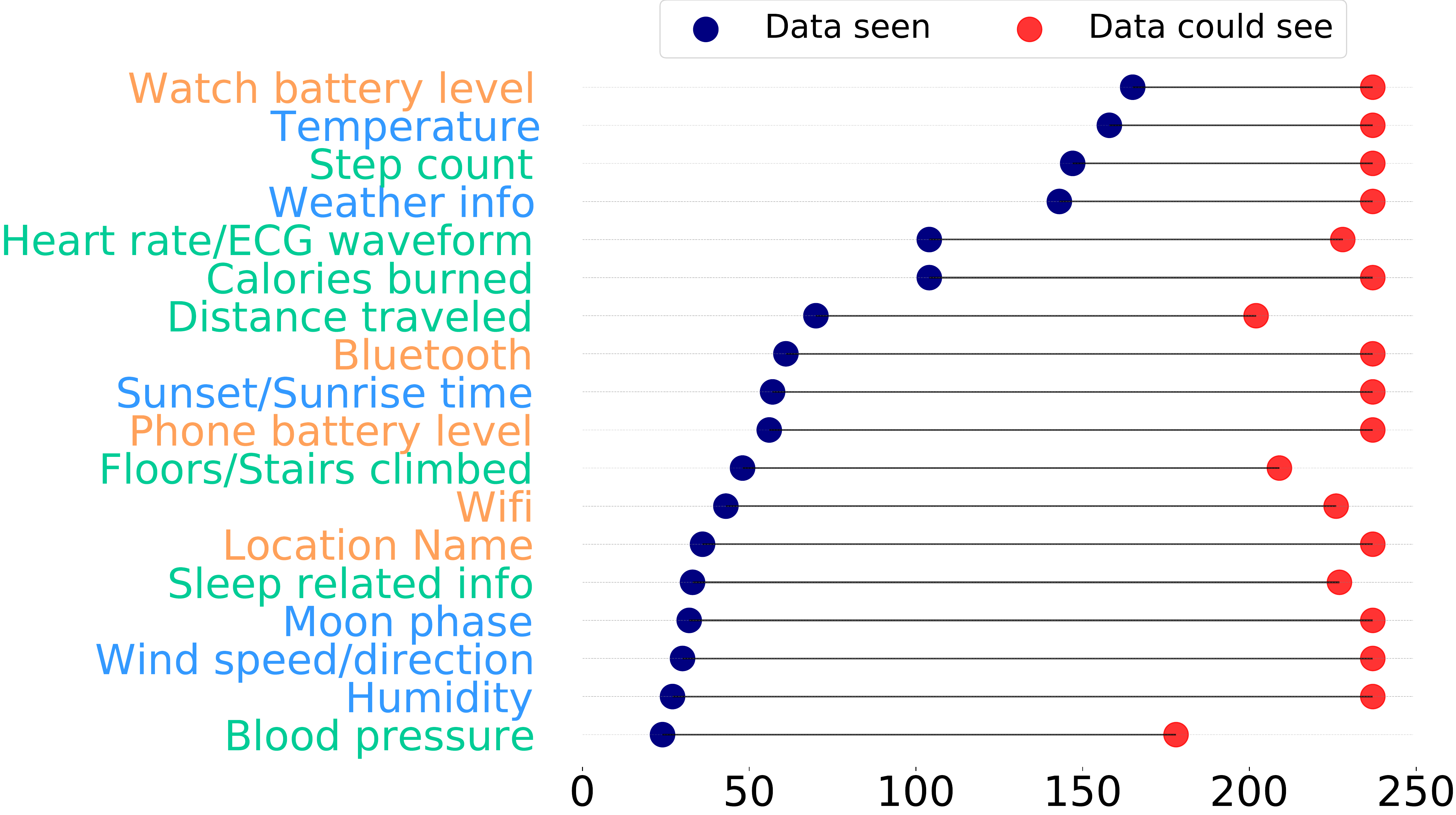}
  \vspace{-7mm}
  \caption{Difference between \# of watches that tracked each data type and how many participants actually saw it on their watch face.}
  \vspace{-3mm}
  \label{fig:datatCouldSeeVsSeen}
\end{figure}

\bpstart{Complementary investigation of device capabilities}
To find untapped opportunities for visual representations, we looked at technical details for the 54 smartwatch models (from the 20 brands) our participants wore. We found that all smartwatches had fitness or activity tracking as a core feature, including measuring and display of body movement, steps, sleep patterns, or dedicated exercise tracking. The smartwatches our participants used also carried a wide variety of sensors \cite{kamivsalic2018sensors}: activity sensors such as accelerometers (53 models) and gyroscopes (46 models); physiological sensors such as heart rate sensors (47 models); and environmental sensors such as barometric altimeters (38 models). Many smartwatches allowed for at least bluetooth (54 models) or wifi (43 models) connectivity. By tracking which types of sensors were available on people's smartwatches, we derived the types of data their watches could track and participants could see on their watch faces (\autoref{fig:datatCouldSeeVsSeen}). 

There naturally is a mismatch between what our participants  could see and what they did see: watch faces do not show all available data. Nevertheless, this mismatch varies. For example, from \health{health \& fitness} data that almost all devices track, roughly 62.03\% of participants see \emph{step counts}, but this percentage is less when it comes to \emph{heart rate} (45.61\%), or \emph{calories burned} (43.88\%), and drops drastically for \emph{distance traveled} (34.65\%), \emph{floors count} (22.97\%), \emph{sleep}, and \emph{blood pressure} (13.48\%). This list of commonly tracked data that is under-represented can serve as a starting point for visualization designers. \changes{For example, in past work \cite{aravind:hal-02337783} we found that smartwatch wearers would have liked to see sleep data but a display on their fitness tracker was not available to them.} 



\section{Discussion and Future Work} 
\label{sec:discussion}

It is challenging to determine a \textit{right} vocabulary for wide-audience surveys. In our case, while we found few examples of icon only displays, participants often reported this type of representation. One possibility for these responses might be confusion about what constitutes ``data.'' In the survey instructions, we informed participants that we only cared about information in the form of numbers or categories, such as step count (numerical) or weather condition (categorical). We also asked participants not to consider graphics such as settings, calendar, or music app icons because they do not represent numerical or categorical information; and gave examples of graphics we cared and did not care about. Yet, participants might not have read the instructions carefully and included responses about graphical icons that do not change based on data. A second possibility for the larger frequency of \icon\ responses might be attributed to typical \icontext\ displays that due to missing or currently inaccessible data result in an icon-only representation (e.g., \inlinevis{-2pt}{1.2em}{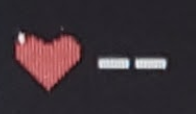}: heart icon with currently blank text). For our analysis reported in \autoref{fig:datatCouldSeeVsSeen} we had to sometimes infer based on sensors whether a certain derived value such as \emph{calories burned} would be available on a watch. The supplementary material makes our inferences transparent.

A wide variety of data types is available for our participants' watch faces. The list of frequently presented data types provides 
starting points for creating visual representations that could be valuable to a broad range of viewers. In addition, when designing perceptual studies in the future, it might be useful to take into account participants' familiarity with this data type.  

Our participants had five data items on average on their watch faces. As five is a relatively large number for a small smartwatch display, an open research question is how to help people cope with such a dense data display. 
Given our analysis of common co-occurrences (especially within the categories) (\autoref{fig:teaser}-right), it may be useful to consider combining them into joint representations. 

Our survey results indicate that visualizations are still not as common as other representations such as text, even though they can be used to represent some of the most commonly displayed data (e.g., \emph{step counts} and \emph{battery levels}). 
Our online search of technical capabilities of smartwatches also indicates that much of the data tracked wearers do not see. This includes some \health{health \& fitness} data that most devices track (e.g., \emph{calories, distance, sleep} and \emph{blood pressure} data).
Whether these are \changes{explicit customization choices due to specific tasks they want to carry out,} or due to a choice the default displays promote for the smartwatch face, remains an open question. Further research needs to investigate representation choices, to determine if the wider adoption of visualizations is a question of preference, tasks, a lack of exposure, and if it requires us to rethink visual encodings for smartwatches. \changes{In addition, future work needs to establish at which level of granularity information should be displayed. For example, are exact wind speeds important or are broad categories (stormy, light breeze, no wind) enough; presentation types would change based on this decision.}

In summary, our work contributes to the understanding of the current real-world use of representation types on smartwatches and additional findings that can inform and inspire the visualization community to pursue smartwatch visualization. 
\acknowledgments{We thank the participants of our survey, and Natkamon Tovanich \& Pierre Dragicevic for their help with the data analysis. The work was funded in part by an ANR grant ANR-18-CE92-0059-01.}

\bibliographystyle{abbrv-doi}

\bibliography{bibliography}

\begin{thebibliography}{10}

\bibitem{8374228}
M.~{Al-Sharrah}, A.~Salman, and I.~{Ahmad}.
\newblock Watch your smartwatch.
\newblock In {\em International Conference on Computing Sciences and
  Engineering (ICCSE)}, pp. 1:1--1:5. IEEE, 2018. doi: {{%
10\hspace{.1pt}\discretionary{.}{%
}{.}\hspace{.4pt}1109\discretionary{/}{%
}{/}ICCSE1\hspace{.1pt}\discretionary{.}{%
}{.}\hspace{.4pt}2018\hspace{.1pt}\discretionary{.}{%
}{.}\hspace{.4pt}8374228}}


\bibitem{10.1145/3154862.3154879}
F.~Amini, K.~Hasan, A.~Bunt, and P.~Irani.
\newblock Data representations for in-situ exploration of health and fitness
  data.
\newblock In {\em Proceedings of the Conference on Pervasive Computing
  Technologies for Healthcare (PervasiveHealth)}, pp. 163--172. ACM, New York,
  NY, USA, 2017. doi: {{%
10\hspace{.1pt}\discretionary{.}{%
}{.}\hspace{.4pt}1145\discretionary{/}{%
}{/}3154862\hspace{.1pt}\discretionary{.}{%
}{.}\hspace{.4pt}3154879}}


\bibitem{Apple:2020}
{Apple Inc. 2015}.
\newblock Apple watch user guide -- version 1.0.
\newblock
  \url{https://manuals.info.apple.com/MANUALS/1000/MA1708/en_US/apple_watch_user_guide.pdf}.
\newblock Last visited: July, 2020.

\bibitem{aravind:hal-02337783}
R.~Aravind, T.~Blascheck, and P.~Isenberg.
\newblock {A Survey on Sleep Visualizations for Fitness Trackers}.
\newblock {Posters of the European Conference on Visualization (EuroVis)},
  2019.
\newblock Poster.

\bibitem{8443125}
T.~{Blascheck}, L.~{Besan\c{c}on}, A.~{Bezerianos}, B.~{Lee}, and
  P.~{Isenberg}.
\newblock Glanceable visualization: Studies of data comparison performance on
  smartwatches.
\newblock {\em IEEE Transactions on Visualization and Computer Graphics},
  25(1):630--640, 2019. doi: {{%
10\hspace{.1pt}\discretionary{.}{%
}{.}\hspace{.4pt}1109\discretionary{/}{%
}{/}TVCG\hspace{.1pt}\discretionary{.}{%
}{.}\hspace{.4pt}2018\hspace{.1pt}\discretionary{.}{%
}{.}\hspace{.4pt}2865142}}


\bibitem{Cecchinato:2017:UESR}
M.~E. Cecchinato, A.~L. Cox, and J.~Bird.
\newblock Always on(line)? user experience of smartwatches and their role
  within multi-device ecologies.
\newblock In {\em Proceedings of the Conference on Human Factors in Computing
  Systems (CHI)}, pp. 3557--3568. ACM, New York, NY, USA, 2017. doi: {{%
10\hspace{.1pt}\discretionary{.}{%
}{.}\hspace{.4pt}1145\discretionary{/}{%
}{/}3025453\hspace{.1pt}\discretionary{.}{%
}{.}\hspace{.4pt}3025538}}


\bibitem{7457170}
J.~{Chauhan}, S.~{Seneviratne}, M.~A. {Kaafar}, A.~{Mahanti}, and
  A.~{Seneviratne}.
\newblock Characterization of early smartwatch apps.
\newblock In {\em Proceedings of the Conference on Pervasive Computing and
  Communication Workshops (PerCom Workshops)}, pp. 1--6, 2016. doi: {{%
10\hspace{.1pt}\discretionary{.}{%
}{.}\hspace{.4pt}1109\discretionary{/}{%
}{/}PERCOMW\hspace{.1pt}\discretionary{.}{%
}{.}\hspace{.4pt}2016\hspace{.1pt}\discretionary{.}{%
}{.}\hspace{.4pt}7457170}}


\bibitem{8880821}
N.~S. {Erdem}, C.~{Ersoy}, and C.~{Tunca}.
\newblock Gait analysis using smartwatches.
\newblock In {\em Proceedings of the Symposium on Personal, Indoor and Mobile
  Radio Communications (PIMRC Workshops)}, pp. 1--6, 2019. doi: {{%
10\hspace{.1pt}\discretionary{.}{%
}{.}\hspace{.4pt}1109\discretionary{/}{%
}{/}PIMRCW\hspace{.1pt}\discretionary{.}{%
}{.}\hspace{.4pt}2019\hspace{.1pt}\discretionary{.}{%
}{.}\hspace{.4pt}8880821}}


\bibitem{10.1145/2971648.2971754}
R.~Gouveia, F.~Pereira, E.~Karapanos, S.~A. Munson, and M.~Hassenzahl.
\newblock Exploring the design space of glanceable feedback for physical
  activity trackers.
\newblock In {\em Proceedings of the Conference on Pervasive and Ubiquitous
  Computing}, pp. 144--155. ACM, New York, NY, USA, 2016. doi: {{%
10\hspace{.1pt}\discretionary{.}{%
}{.}\hspace{.4pt}1145\discretionary{/}{%
}{/}2971648\hspace{.1pt}\discretionary{.}{%
}{.}\hspace{.4pt}2971754}}


\bibitem{7299348}
H.~{Kalantarian}, N.~{Alshurafa}, E.~{Nemati}, T.~{Le}, and M.~{Sarrafzadeh}.
\newblock A smartwatch-based medication adherence system.
\newblock In {\em Proceedings of the Conference on Wearable and Implantable
  Body Sensor Networks (BSN)}, pp. 1--6, 2015. doi: {{%
10\hspace{.1pt}\discretionary{.}{%
}{.}\hspace{.4pt}1109\discretionary{/}{%
}{/}BSN\hspace{.1pt}\discretionary{.}{%
}{.}\hspace{.4pt}2015\hspace{.1pt}\discretionary{.}{%
}{.}\hspace{.4pt}7299348}}


\bibitem{kamivsalic2018sensors}
A.~Kami{\v{s}}ali{\'c}, I.~Fister, M.~Turkanovi{\'c}, and S.~Karakati{\v{c}}.
\newblock Sensors and functionalities of non-invasive wrist-wearable devices: A
  review.
\newblock {\em Sensors}, 18(6):1714, 2018. doi: {{%
10\hspace{.1pt}\discretionary{.}{%
}{.}\hspace{.4pt}3390\discretionary{/}{%
}{/}s18061714}}


\bibitem{10.1145/3313831.3376199}
K.~Klamka, T.~Horak, and R.~Dachselt.
\newblock Watch+strap: Extending smartwatches with interactive strapdisplays.
\newblock In {\em Proceedings of the Conference on Human Factors in Computing
  Systems (CHI)}, pp. 1--15. ACM, New York, NY, USA, 2020. doi: {{%
10\hspace{.1pt}\discretionary{.}{%
}{.}\hspace{.4pt}1145\discretionary{/}{%
}{/}3313831\hspace{.1pt}\discretionary{.}{%
}{.}\hspace{.4pt}3376199}}


\bibitem{Facer:2020}
{Little Labs, Inc.}
\newblock Facer - thousands of free watch faces for apple watch, samsung gear
  s3, huawei watch, and more.
\newblock \url{https://www.facer.io/}.
\newblock Last visited: July, 2020.

\bibitem{10.1145/2935334.2935344}
K.~Lyons.
\newblock Visual parameters impacting reaction times on smartwatches.
\newblock In {\em Proceedings of Conference on Human-Computer Interaction with
  Mobile Devices and Services (MobileHCI)}, pp. 190--194. ACM, New York, NY,
  USA, 2016. doi: {{%
10\hspace{.1pt}\discretionary{.}{%
}{.}\hspace{.4pt}1145\discretionary{/}{%
}{/}2935334\hspace{.1pt}\discretionary{.}{%
}{.}\hspace{.4pt}2935344}}


\bibitem{10.1145/3341162.3346276}
M.~Maritsch, C.~B\'{e}rub\'{e}, M.~Kraus, V.~Lehmann, T.~Z\"{u}ger,
  S.~Feuerriegel, T.~Kowatsch, and F.~Wortmann.
\newblock Improving heart rate variability measurements from consumer
  smartwatches with machine learning.
\newblock In {\em Adjunct Proceedings of the Conference on Pervasive and
  Ubiquitous Computing and Proceedings of the Symposium on Wearable Computers},
  pp. 934--938. ACM, New York, NY, USA, 2019. doi: {{%
10\hspace{.1pt}\discretionary{.}{%
}{.}\hspace{.4pt}1145\discretionary{/}{%
}{/}3341162\hspace{.1pt}\discretionary{.}{%
}{.}\hspace{.4pt}3346276}}


\bibitem{10.1145/3025453.3025993}
D.~McMillan, B.~Brown, A.~Lampinen, M.~McGregor, E.~Hoggan, and S.~Pizza.
\newblock Situating wearables: Smartwatch use in context.
\newblock In {\em Proceedings of the Conference on Human Factors in Computing
  Systems (CHI)}, pp. 3582--3594. ACM, New York, NY, USA, 2017. doi: {{%
10\hspace{.1pt}\discretionary{.}{%
}{.}\hspace{.4pt}1145\discretionary{/}{%
}{/}3025453\hspace{.1pt}\discretionary{.}{%
}{.}\hspace{.4pt}3025993}}


\bibitem{Neis:2016}
S.~M. Neis and M.~I. Blackstun.
\newblock Feasibility analysis of wearables for use by airline crew.
\newblock In {\em 2016 IEEE/AIAA 35th Digital Avionics Systems Conference
  (DASC)}, pp. 1--9, 2016. doi: {{%
10\hspace{.1pt}\discretionary{.}{%
}{.}\hspace{.4pt}1109\discretionary{/}{%
}{/}DASC\hspace{.1pt}\discretionary{.}{%
}{.}\hspace{.4pt}2016\hspace{.1pt}\discretionary{.}{%
}{.}\hspace{.4pt}7778023}}


\bibitem{PMID:30741218}
A.~Neshati, Y.~Sakamoto, and P.~Irani.
\newblock Challenges in displaying health data on small smartwatch screens.
\newblock {\em Studies in Health Technology and Informatics}, 257:325--332,
  2019. doi: {{%
10\hspace{.1pt}\discretionary{.}{%
}{.}\hspace{.4pt}3233\discretionary{/}{%
}{/}978\discretionary{%
}{-}{-}1\discretionary{%
}{-}{-}61499\discretionary{%
}{-}{-}951\discretionary{%
}{-}{-}5\discretionary{%
}{-}{-}325}}


\bibitem{Neshati:2019:10.20380/GI2019.23}
A.~Neshati, Y.~Sakamoto, L.~C. Leboe-McGowan, J.~Leboe-McGowan, M.~Serrano, and
  P.~Irani.
\newblock G-sparks: Glanceable sparklines on smartwatches.
\newblock In {\em Proceedings of Graphics Interface (GI)}, pp. 23--1. Canadian
  Information Processing Society, 2019. doi: {{%
10\hspace{.1pt}\discretionary{.}{%
}{.}\hspace{.4pt}20380\discretionary{/}{%
}{/}GI2019\hspace{.1pt}\discretionary{.}{%
}{.}\hspace{.4pt}23}}


\bibitem{10.1145/2858036.2858522}
S.~Pizza, B.~Brown, D.~McMillan, and A.~Lampinen.
\newblock Smartwatch in vivo.
\newblock In {\em Proceedings of the Conference on Human Factors in Computing
  Systems (CHI)}, pp. 5456--5469. ACM, New York, NY, USA, 2016. doi: {{%
10\hspace{.1pt}\discretionary{.}{%
}{.}\hspace{.4pt}1145\discretionary{/}{%
}{/}2858036\hspace{.1pt}\discretionary{.}{%
}{.}\hspace{.4pt}2858522}}


\bibitem{10.1145/2851581.2892493}
R.~Quintana, C.~Quintana, C.~Madeira, and J.~D. Slotta.
\newblock Keeping watch: Exploring wearable technology designs for k-12
  teachers.
\newblock In {\em Extended Abstracts of the Conference on Human Factors in
  Computing Systems (CHI)}, pp. 2272--2278. ACM, New York, NY, USA, 2016. doi:
  {{%
10\hspace{.1pt}\discretionary{.}{%
}{.}\hspace{.4pt}1145\discretionary{/}{%
}{/}2851581\hspace{.1pt}\discretionary{.}{%
}{.}\hspace{.4pt}2892493}}


\bibitem{10.1145/3123024.3125614}
J.~C. Quiroz, M.~H. Yong, and E.~Geangu.
\newblock Emotion-recognition using smart watch accelerometer data: Preliminary
  findings.
\newblock In {\em Proceedings of the Conference on Pervasive and Ubiquitous
  Computing and Proceedings of the Symposium on Wearable Computers}, pp.
  805--812. ACM, New York, NY, USA, 2017. doi: {{%
10\hspace{.1pt}\discretionary{.}{%
}{.}\hspace{.4pt}1145\discretionary{/}{%
}{/}3123024\hspace{.1pt}\discretionary{.}{%
}{.}\hspace{.4pt}3125614}}


\bibitem{RAM:2020}
{Research And Markets}.
\newblock Smartwatch market - growth, trends, forecasts (2020 - 2025).
\newblock
  \url{https://www.researchandmarkets.com/reports/4591978/smartwatch-market-growth-trends-forecasts}.
\newblock Last visited: July, 2020.

\bibitem{Schirra:2015:UEUCSW}
S.~Schirra and F.~R. Bentley.
\newblock ``it’s kind of like an extra screen for my phone'': Understanding
  everyday uses of consumer smart watches.
\newblock In {\em Extended Abstracts of the Conference on Human Factors in
  Computing Systems (CHI)}, pp. 2151--2156. ACM, New York, NY, USA, 2015. doi:
  {{%
10\hspace{.1pt}\discretionary{.}{%
}{.}\hspace{.4pt}1145\discretionary{/}{%
}{/}2702613\hspace{.1pt}\discretionary{.}{%
}{.}\hspace{.4pt}2732931}}


\bibitem{10.1145/3341162.3344831}
P.~Siirtola.
\newblock Continuous stress detection using the sensors of commercial
  smartwatch.
\newblock In {\em Adjunct Proceedings of the Conference on Pervasive and
  Ubiquitous Computing and Proceedings of the Symposium on Wearable Computers},
  pp. 1198--1201. ACM, New York, NY, USA, 2019. doi: {{%
10\hspace{.1pt}\discretionary{.}{%
}{.}\hspace{.4pt}1145\discretionary{/}{%
}{/}3341162\hspace{.1pt}\discretionary{.}{%
}{.}\hspace{.4pt}3344831}}


\bibitem{10.5555/3400306.3400354}
S.~Stankoski, N.~Re\v{s}\v{c}i\v{c}, G.~Me\v{z}i\v{c}, and M.~Lu\v{s}trek.
\newblock Real-time eating detection using a smartwatch.
\newblock In {\em Proceedings of the Conference on Embedded Wireless Systems
  and Networks (EWSN)}, pp. 247--252. Junction Publishing, USA, 2020.

\bibitem{e2da2cc89b474b6ebbf3817b4bb025c6}
A.~Suciu and J.~Larsen.
\newblock Active self-tracking and visualization of subjective experience using
  {VAS} and time spirals on a smartwatch.
\newblock In {\em Proceedings of the Data Visualization on Mobile Devices
  Workshop held at the ACM Conference on Human Factor in Computing Systems
  (CHI)}, 2018.

\bibitem{10.1145/3025453.3025817}
A.~Visuri, Z.~Sarsenbayeva, N.~van Berkel, J.~Goncalves, R.~Rawassizadeh,
  V.~Kostakos, and D.~Ferreira.
\newblock Quantifying sources and types of smartwatch usage sessions.
\newblock In {\em Proceedings of the Conference on Human Factors in Computing
  Systems (CHI)}, pp. 3569--3581. ACM, New York, NY, USA, 2017. doi: {{%
10\hspace{.1pt}\discretionary{.}{%
}{.}\hspace{.4pt}1145\discretionary{/}{%
}{/}3025453\hspace{.1pt}\discretionary{.}{%
}{.}\hspace{.4pt}3025817}}


\bibitem{Zhang:2019:EnergyInefficienciesWatchFaces}
H.~Zhang, H.~Wu, and A.~Rountev.
\newblock Detection of energy inefficiencies in android wear watch faces.
\newblock In {\em Proceedings of the Meeting on European Software Engineering
  Conference and Symposium on the Foundations of Software Engineering}, pp.
  691--702. ACM, New York, NY, USA, 2018. doi: {{%
10\hspace{.1pt}\discretionary{.}{%
}{.}\hspace{.4pt}1145\discretionary{/}{%
}{/}3236024\hspace{.1pt}\discretionary{.}{%
}{.}\hspace{.4pt}3236073}}


\end{thebibliography}
\end{document}